\documentclass[aps,pra,showpacs,floatfix]{revtex4}
\usepackage{multirow}
\usepackage{graphics}
\usepackage{epsfig}
\usepackage{subfigure}
\usepackage{amsmath,amsfonts,amssymb,graphicx}
\begin{document}
\title{A general scheme for multiparty controlled quantum teleportation of an arbitrary N-particle state}
\author{Jun Ye$^{\dag}$}
\author{Yue Li$^{\dag,\S}$}\email{liyue@smail.hust.edu.cn}
\author{Yu Liu$^{\dag}$}
\author{Sha Hua$^{\P}$}
\affiliation{$^{\dag}$Department of Electronics \& Information
Engineering, Huazhong University of Science \& Technology, Wuhan
430074, China\\
$^{\S}$School of Computer Science \& Technology, Huazhong University
of Science \& Technology, Wuhan 430074, China\\
$^{\P}$Department of Electrical \& Computer Engineering, Polytechnic
University, New York 11373, USA}
\date{March 27, 2008}
\begin{abstract}
There is much interest in the multiparty quantum communications
where quantum teleportation using high dimensional entangled quantum
channel is one of the promising tools. In this paper, we propose a
more general scheme for M-party controlled teleportation of an
arbitrary N-particle quantum state using N-1 identical
Einstein-Podolsky-Rosen pairs and one (M+2)-particle
Greenberger-Horne-Zeilinger state together as quantum channel. Based
on which a 2-party controlled teleportation of an arbitrary
3-particle state is tested with our scheme as an example.
\end{abstract}
 \maketitle
\section{Introduction} %introduction
Quantum teleportation (QT) \cite{1} is considered as one of the most
successful inventions in the quantum information research \cite{0},
it achieves the destination of quantum state transfer in a different
way via prior shared entanglement as the quantum channel as well as
the local operation and classical communication (LOCC), letting an
arbitrary qubit collapses from the sender and ``reborns" at the
receiver's side without distant qubit traveling. It plays
significant roles both in the theoretical and experimental fields of
quantum information processing. Since the first theoretical protocol
for teleporting an unknown qubit in 1993 \cite{1}, this technology
has been extensively
studied~\cite{2,3,4,5,6,7,8,9,10,11,12,13,14,15,16,17,18,19,20,21,22,221,222,23,24,25}:
QT of a bipartite Einstein-Podolsky-Rosen (EPR) pair~\cite{28} were
presented \cite{9,10,11,12,13}; schemes which teleport a tripartite
Greenberger-Horne-Zeilinger (GHZ) state \cite{26} as well as
tripartite W state \cite{27} were also investigated
\cite{14,15,16,17}.

Later, quantum communications were brought into a multiparty based
environment and the QT using high dimensional entangled quantum
channel is considered as one of the promising tools for this task.
This definitely calls for general QT schemes in higher dimensional
space. Recently, QT of an arbitrary N-particle state ($N\geq 1$) via
$N$ EPR states as quantum channel was proposed~\cite{18} and a
scheme for teleporting an unknown N-particle W state via
entanglement swapping~\cite{ZZHE93} was also introduced~\cite{19};
based on multiparticle entanglement theory in Ref.~\cite{13}, a
general scheme for teleporting an arbitrary N-particle state via
genuine N-particle entangled quantum channel was given in
Ref.~\cite{20}.

More recently, combining with the idea of controlled communication,
controlled quantum teleportation (CQT), which allows QT being
performed under a third party's permission, has been paid much
attention and developed into the situation where more than one
controller are needed~\cite{21,22,221,222,23,24,25}: CQT for
teleporting a single-particle state have been
studied~\cite{21,22,221,222} and multiparticle quantum state from
Alice to Bob under the control of many networked agents was also
provided~\cite{23}; A method for symmetric multiparty controlled
quantum teleportation (MCQT) of an arbitrary bipartite entanglement
using two GHZ states was presented \cite{24} and MCQT of a
N-particle W state via N-1 EPR pairs and one (M+2)-particle GHZ
state was proposed \cite{25}. In this paper we completely summarize
the schemes presented and propose a more general scheme for M-party
MCQT of an \emph{arbitrary} N-particle state via N-1 EPR pairs and
an (M+2)-particle GHZ state as quantum channel.

\section{The MCQT of an arbitrary N-particle state}
\subsection{Assumptions}
\label{sec:2} Suppose the sender Alice wants to teleport an unknown
N-particle state to the distant receiver Bob, and she is only able
to do so as long as all $M$ agents $Charlie_{1}, Charlie_{2}, ... ,
Charlie_{M}$ permit, here the original arbitrary unknown state reads
\begin{equation}
\begin{split}
|\phi\rangle_{A_{1}A_{2}...A_{N}}=&(x_{1}|000...00\rangle+x_{2}|000...01\rangle+...+x_{2^{N}}|111...11\rangle)_{A_{1}A_{2}...A_{N}}\\
=&\sum^{2^N}_{k=1} x_{k}\prod_{i=1}^{N}
|\delta_{kA_{i}}\rangle_{A_{i}},
\end{split}
\end{equation}
where $\sum^{2^N}_{k=1}|x_{k}|^{2}=1, x_{k}\neq
0,\delta_{kA_{i}}\in\{0, 1\} (i=1, 2, ... , N)$. To achieve the QT
task, Alice and Bob as well as the M agents take use of the quantum
channel which is made of ($N-1$) identical EPR pairs and an
(M+2)-particle GHZ state and follow our MCQT protocol. We assume
each EPR pair used in the channel are in the state
\begin{equation}
|\phi^{+}\rangle=\frac{1}{\sqrt{2}}(|00\rangle+|11\rangle),
\end{equation}
and cases for other EPR pairs: $|\phi^{-}\rangle, |\psi^{+}\rangle$
and $|\psi^{-}\rangle$ are studied in the appendix.
\subsection{The scheme}
We detail the scheme in the following steps:
\begin{itemize}
\item[1.] Alice prepares $(N-1)$ EPR pairs, taking the joint state
\begin{equation}
    \begin{split}
    |\phi\rangle_{DB}&=\prod_{i=1}^{N-1}|\phi^{+}\rangle_{D_{i}B_{i}}\\
    &=\prod_{i=1}^{N-1}\frac{1}{\sqrt{2}}(|00\rangle+|11\rangle)_{D_{i}B_{i}}
    \end{split}
    \end{equation}
with one (M+2)-particle GHZ state
\begin{equation}
|\phi^{+}\rangle_{A_{N}B_{N}C_{1}...C_{M}}
=\frac{1}{\sqrt{2}}(|000...0\rangle+|111...1\rangle)_{A_{N}B_{N}C_{1}...C_{M}}
    \end{equation}
together as quantum channel.
    \item[2.] Alice sends the (N-1) particles $(B_{1},B_{2}, ... ,B_{N-1})$ from the state $|\phi\rangle_{DB}$ and the GHZ particle $B_{N}$ to Bob, she sends the $M$ GHZ particles  $(C_{1},C_{2},...,C_{M})$ to $M$ agents ($Charlie_{1}, Charlie_{2},...,Charlie_{M}$) respectively while keeping the other $(N-1)$ EPR particles $(D_{1},D_{2},...,D_{N-1})$ and one GHZ particle $A_{N}$ to herself. Therefore, including the unknown state $|\phi\rangle_{A_{1}A_{2}...A_{N}}$, the whole system state reads
\begin{equation}
\begin{split}
|\psi_{s}\rangle&=|\phi_{A_{1}A_{2}...A_{N}}\rangle\otimes|\phi\rangle_{DB}\otimes|\phi^{+}\rangle_{D_{N}B_{N}C_{1}...C_{M}}\\
&=|\phi_{A_{1}A_{2}...A_{N}}\rangle\otimes(|\phi^{+}\rangle_{D_{1}B_{1}}\otimes|\phi^{+}\rangle_{D_{2}B_{2}}\otimes ... \otimes|\phi^{+}\rangle_{D_{N-1}B_{N-1}})\otimes|\phi^{+}\rangle_{D_{N}B_{N}C_{1}...C_{M}}\\
&=\sum^{2^{N}}_{k=1}x_{k}\prod^{N}_{i=1}|\delta_{kA_{i}}\rangle_{A_{i}}\otimes\prod^{N-1}_{i=1}\frac{1}{\sqrt{2}}(|00\rangle+|11\rangle)_{D_{i}B_{i}}\otimes\frac{1}{\sqrt{2}}(|000...0\rangle+|111...1\rangle)_{A_{N}B_{N}C_{1}...C_{M}}.
\end{split}
\end{equation}
\item[3.] Alice performs the Bell-joint measurements on pairs $(A_{i}, D_{i}), (i=1, 2, ..., N-1)$ separately, obtaining result $|\psi\rangle_{A_{i}D_{i}}$. Then she broadcasts these results via classical channel, according to which Bob correspondingly performs a series of single-particle unitary transformations $U_{i}$ on particles $B_{i}, (i=1, 2, ... , N-1)$, see Tab.~\ref{tab:1}.
\begin{table}[htbp]
\caption{Correlations between Alice's measurement results and Bob's
unitary transformation} \label{tab:1}
\begin{center}
\begin{tabular}{c    c} %% this creates two columns
\hline \hline
\rule[-1ex]{0pt}{3.5ex}  Alice's measurement results & Bob's unitary transformation\\
\rule[-1ex]{0pt}{3.5ex}  $|\psi\rangle_{A_{i}D_{i}}$ & $U_{i}$\\
\hline \hline
\rule[-1ex]{0pt}{3.5ex}  $|\phi^{+}\rangle$ & $I$ \\
\rule[-1ex]{0pt}{3.5ex}  $|\phi^{-}\rangle$ & $\sigma_{z}$\\
\rule[-1ex]{0pt}{3.5ex}  $|\psi^{+}\rangle$ & $\sigma_{x}$\\
\rule[-1ex]{0pt}{3.5ex}  $|\psi^{-}\rangle$ & $i\sigma_{y}$\\
\hline
\end{tabular}
\end{center}
\end{table}
\item[4.] Alice performs Bell-joint measurements on the pairs $(A_{N}, D_{N})$ obtaining state $|\psi\rangle_{A_{N}D_{N}}$ and also publishes her results. According to the results, this time, Bob performs a corresponding unitary transformation $U_{N}$ on the particle $B_{N}$, then the $M$ agents perform a series of single-particle unitary transformations $U_{C_{1}},U_{C_{2}},...,U_{C_{M}}$ on the particles $C_{1},C_{2},...,C_{M}$ respectively, see Tab.~\ref{tab:2}.
\begin{table}[h]
\caption{Correlations among Alice's measurement results, Bob and
$Charlie_{i}$'s unitary transformations}
    \label{tab:2}
    \begin{center}
    \begin{tabular}{c    c    c} %% this creates two columns
    \hline
    \hline
    \rule[-1ex]{0pt}{3.5ex}  Alice's measurement results & Bob's unitary transformation & $Charlie_{i}$'s unitary transformation\\
    \rule[-1ex]{0pt}{3.5ex}  $|\psi\rangle_{A_{N}D_{N}}$ & $U_{N}$ & $U_{C_{i}}$\\
    \hline \hline
    \rule[-1ex]{0pt}{3.5ex}  $|\phi^{+}\rangle$ & $I$ & $I$ \\
    \rule[-1ex]{0pt}{3.5ex}  $|\phi^{-}\rangle$ & $\sigma_{z}$ & $I$\\
    \rule[-1ex]{0pt}{3.5ex}  $|\psi^{+}\rangle$ & $\sigma_{x}$ & $\sigma_{x}$\\
    \rule[-1ex]{0pt}{3.5ex}  $|\psi^{-}\rangle$ & $i\sigma_{y}$ & $i\sigma_{y}$\\
    \hline
    \end{tabular}
    \end{center}
    \end{table}
After the above steps, the state of the whole system becomes
\begin{equation}
\label{eq:1}
\begin{split}
|\psi_{s}\rangle_{1}&=U_{C_{M}}...U_{C_{1}}U_{N}U_{N-1}...U_{2}U_{1}\
_{A_{N-1}D_{N-1}}\langle
\psi| ... \ _{A_{2}D_{2}}\langle\psi|\ _{A_{1}D_{1}}\langle\psi|\psi_{s}\rangle\\
&=\sum^{2^{N}}_{k=1}x_{k}[(\sum^{N-1}_{i=1}U_{i}\
_{A_{i}D_{i}}\langle\psi|\delta_{kA_{i}}\rangle_{A_{i}}|\phi^{+}\rangle_{D_{i}B_{i}})
\otimes(U_{N}U_{C_{M}} ... U_{C_{1}}\otimes
_{A_{N}D_{N}}\langle\psi|\delta_{kA_{i}}\rangle_{A_{N}}|\phi^{+}\rangle_{D_{N}B_{N}C_{1}
... C_{M}})],
\end{split}
\end{equation}
where $\ _{A_{i}D_{i}}\langle\psi|$ is the Bell-joint measurement on
pair $(A_{i}, D_{i})$.

Now we analyze the whole system in Eq.~\ref{eq:1}. For simplicity,
define
\begin{equation}
\label{eq:2} |\beta_{B_{i}}\rangle=U_{i}\
_{A_{i}D_{i}}\langle\psi|\delta_{kA_{i}}\rangle_{A_{i}}|\phi^{+}\rangle_{D_{i}B_{i}},
\end{equation}
\begin{equation}
\label{eq:10} |\beta_{B_{N}C_{1} ...
C_{M}}\rangle=U_{N}U_{C_{M}}...U_{C_{1}}\otimes\
_{A_{N}D_{N}}\langle\psi|\delta_{kA_{i}}\rangle_{A_{N}}|\phi^{+}\rangle_{D_{N}B_{N}C_{1}...C_{M}}.
\end{equation}
From Eq.~\ref{eq:2} and Tab.~\ref{tab:1}, we have
\begin{equation}
\begin{split}
|\beta_{B_{i}}\rangle&=
\begin{cases}
I_{B_{i}}(\
_{A_{i}}\langle0|\delta_{kA_{i}}\rangle_{A_{i}}|0\rangle_{B_{i}}+\
_{A_{i}} \langle1|\delta_{kA_{i}}\rangle_{A_{i}}|1\rangle_{B_{i}})&
\text{for
$|\psi\rangle_{A_{i}D_{i}}=|\phi^{+}\rangle$}\\
(\sigma_{z})_{B_{i}}(\
_{A_{i}}\langle0|\delta_{kA_{i}}\rangle_{A_{i}}|0\rangle_{B_{i}}-\
_{A_{i}} \langle1|\delta_{kA_{i}}\rangle_{A_{i}}|1\rangle_{B_{i}})&
\text{for
$|\psi\rangle_{A_{i}D_{i}}=|\phi^{-}\rangle$}\\
(\delta_{x})_{B_{i}}(\
_{A_{i}}\langle0|\delta_{kA_{i}}\rangle_{A_{i}}|1\rangle_{B_{i}}+\
_{A_{i}} \langle1|\delta_{kA_{i}}\rangle_{A_{i}}|0\rangle_{B_{i}})&
\text{for
$|\psi\rangle_{A_{i}D_{i}}=|\psi^{+}\rangle$}\\
(i\delta_{y})_{B_{i}}(\
_{A_{i}}\langle0|\delta_{kA_{i}}\rangle_{A_{i}}|1\rangle_{B_{i}}-\
_{A_{i}} \langle1|\delta_{kA_{i}}\rangle_{A_{i}}|0\rangle_{B_{i}})&
\text{for $|\psi\rangle_{A_{i}D_{i}}=|\psi^{-}\rangle$}
\end{cases}
\\
&=\
_{A_{i}}\langle0|\delta_{kA_{i}}\rangle_{A_{i}}|0\rangle_{B_{i}}+\
_{A_{i}}\langle1|\delta_{kA_{i}}\rangle_{A_{i}}|1\rangle_{B_{i}}\\
&=
\begin{cases}
|0\rangle_{B_{i}},|\delta_{kA_{i}}\rangle_{A_{i}}=0&\\
|1\rangle_{B_{i}},|\delta_{kA_{i}}\rangle_{A_{i}}=1&
\end{cases}
\\
&=|\delta_{kA_{i}}\rangle_{B_{i}};
\end{split}
\end{equation}
From Eq.~\ref{eq:10} and Tab.~\ref{tab:2}, we have
\begin{equation}
\begin{split}
|\beta_{B_{N}C_{1}...C_{M}}\rangle&=
\begin{cases}
I_{B_{N}}(\prod_{j=1}^{M}I_{C_{j}})(\
_{A_{N}}\langle0|\delta_{A_{N}}\rangle_{kA_{N}}|000...0\rangle_{B_{N}C_{1}C_{2}...C_{M}}+
\
_{A_{N}}\langle1|\delta_{kA_{N}}\rangle_{A_{N}}|111...1\rangle_{B_{N}C_{1}C_{2}...C_{M}})&
\\\text{for $|\psi\rangle_{A_{N}D_{N}}=|\phi^{+}\rangle$}\\\\
(\sigma_{z})_{B_{N}}(\prod_{j=1}^{M}I_{C_{j}})(\
_{A_{N}}\langle0|\delta_{A_{N}}\rangle_{kA_{N}}|000...0\rangle_{B_{N}C_{1}C_{2}...C_{M}}-
\
_{A_{N}}\langle1|\delta_{kA_{N}}\rangle_{A_{N}}|111...1\rangle_{B_{N}C_{1}C_{2}...C_{M}})&
\\\text{for $|\psi\rangle_{A_{N}D_{N}}=|\phi^{-}\rangle$}\\\\
(\sigma_{x})_{B_{N}}[\prod_{j=1}^{M}(\sigma_{x})_{C_{j}}](\
_{A_{N}}\langle0|\delta_{A_{N}}\rangle_{kA_{N}}|111...1\rangle_{B_{N}C_{1}C_{2}...C_{M}}+
\
_{A_{N}}\langle1|\delta_{kA_{N}}\rangle_{A_{N}}|000...0\rangle_{B_{N}C_{1}C_{2}...C_{M}})&
\\\text{for $|\psi\rangle_{A_{N}D_{N}}=|\psi^{+}\rangle$}\\\\
(i\sigma_{y})_{B_{N}}[\prod_{j=1}^{M}(i\sigma_{y})_{C_{j}}](\
_{A_{N}}\langle0|\delta_{A_{N}}\rangle_{kA_{N}}|111 ...
1\rangle_{B_{N}C_{1}C_{2} ... C_{M}}- \
_{A_{N}}\langle1|\delta_{kA_{N}}\rangle_{A_{N}}|000 ...
0\rangle_{B_{N}C_{1}C_{2} ...C_{M}})&
\\\text{for $|\psi\rangle_{A_{N}D_{N}}=|\psi^{-}\rangle$}
\end{cases}
\\
&=\ _{A_{N}}\langle0|\delta_{kA_{N}}\rangle_{A_{N}}|000 ...
0\rangle_{B_{N}C_{1}C_{2} ... C_{M}}+\ _{A_{N}}\langle1|\delta_{kA_{N}}\rangle_{A_{N}}|111 ... 1\rangle_{B_{N}C_{1}C_{2} ... C_{M}}\\
&=
\begin{cases}
|111...1\rangle_{B_{N}C_{1}C_{2} ... C_{M}},
|\delta_{kA_{N}}\rangle_{A_{N}}=1&\\
|000...0\rangle_{B_{N}C_{1}C_{2} ... C_{M}},
|\delta_{kA_{N}}\rangle_{A_{N}}=0&
\end{cases}
\\
&=|\delta_{A_{N}}\delta_{A_{N}}\delta_{A_{N}} ...
\delta_{A_{N}}\rangle_{B_{N}C_{1}C_{2} ... C_{M}}.
\end{split}
\end{equation}
Therefore, the whole system in Eq.~\ref{eq:1} reads
\begin{equation}
\begin{split}
\label{eq:13}
|\psi_{s}\rangle_{1}&=\sum^{2^{N}}_{k=1}x_{k}(\prod^{N-1}_{i=1}|\delta_{kA_{i}}\rangle_{B_{i}}\otimes|\delta_{A_{N}}\delta_{A_{N}}\delta_{A_{N}}...\delta_{A_{N}}\rangle_{B_{N}C_{1}C_{2}...C_{M}})\\
&=\sum^{2^{N}}_{k=1}x_{k}(|\delta_{kA_{2}}...\delta_{kA_{N-1}}\rangle_{B_{1}B_{2}...B_{N-1}}\otimes|\delta_{kA_{N}}\delta_{kA_{N}}...\delta_{kA_{N}}\rangle_{B_{N}C_{1}C_{2}...C_{M}})\\
&=\sum^{2^{N}}_{k=1}x_{k}(\prod^{N-1}_{i=1}|\delta_{kA_{i}}\rangle_{B_{i}}\otimes|\delta_{kA_{N}}\rangle_{B_{N}}\otimes|\delta_{kA_{N}}\delta_{kA_{N}}...\delta_{kA_{N}}\rangle_{C_{1}C_{2}...C_{M}}).
\end{split}
\end{equation}
\item[5.] The $M$ agents ($Charlie_{1}, Charlie_{2}, ... , Charlie_{M}$)  respectively perform the Hadamard transformations on the particles in their hands. Since $\mathcal{H}|0\rangle=\frac{1}{\sqrt{2}}(|0\rangle+|1\rangle)$ and
$\mathcal{H}|1\rangle=\frac{1}{\sqrt{2}}(|0\rangle-|1\rangle$), the
whole system in Eq.~\ref{eq:13} becomes
\begin{equation}
\label{eq:14}
\begin{split}
|\psi_{s}\rangle_{2}=&(x_{1}+x_{3}+ ... +x_{2^{N-1}})[\prod^{N-1}_{i=1}|\delta_{kA_{i}}\rangle_{B_{i}}\otimes|0\rangle_{B_{N}}\otimes\prod^{M}_{j=1}(|0\rangle+|1\rangle)_{C_{j}}]\\
&+(x_{2}+x_{4}+ ... +x_{2^N})[\prod^{N-1}_{i=1}|\delta_{kA_{i}}\rangle_{B_{i}}\otimes|1\rangle_{B_{N}}\otimes\prod^{M}_{j=1}(|0\rangle-|1\rangle)_{C_{j}}]\\
=&(x_{1}+x_{3}+ ... +x_{2^{N}-1})[\prod^{N-1}_{i=1}|\delta_{kA_{i}}\rangle_{B_{i}}\otimes|0\rangle_{B_{N}}\otimes\sum(|C_{e}\rangle_{C_{1}C_{2}...C_{M}}+|C_{0}\rangle_{C_{1}C_{2}...C_{M}})]\\
&+(x_{2}+x_{4}+ ... +x_{2^N}[\prod^{N-1}_{i=1}|\delta_{kA_{i}}\rangle_{B_{i}}\otimes|1\rangle_{B_{N}}\otimes\sum(|C_{e}\rangle_{C_{1}C_{2}...C_{M}}-|C_{0}\rangle_{C_{1}C_{2}...C_{M}})]\\
=&\sum|C_{e}\rangle_{C_{1}C_{2}...C_{M}}[(x_{1}+x_{3}+ ... +x_{2^{N}-1})\prod^{N-1}_{i=1}|\delta_{kA_{i}}\rangle_{B_{i}}\otimes|0\rangle_{B_{N}}+(x_{2}+x_{4}+ ... +x_{2^{N}})\prod^{N-1}_{i=1}|\delta_{kA_{i}}\rangle_{B_{i}}\otimes|1\rangle_{B_{N}}]\\
&+\sum|C_{0}\rangle_{C_{1}C_{2}...C_{M}}[(x_{1}+x_{3}+ ...
+x_{2^{N}-1})\prod^{N-1}_{i=1}|\delta_{kA_{i}}\rangle_{B_{i}}\otimes|0\rangle_{B_{N}}-(x_{2}+x_{4}+
...
+x_{2^{N}})\prod^{N-1}_{i=1}|\delta_{kA_{i}}\rangle_{B_{i}}\otimes|1\rangle_{B_{N}}].
\end{split}
\end{equation}
\item[6.] To permit Alice and Bob's QT request, each agent measures his particle in basis $\{|0\rangle,|1\rangle\}$, and publishes his result in ``0" or ``1" correspondingly. According to the number of ``1" in the $M$ agents' published measurement results, Bob choose different operations as below to fully recover the original unknown N-particle state of Alice:
\begin{itemize}
    \item If the $M$ agents' measurement results contain odd number of ``1", Bob's N-particle system must be in the state
    \begin{equation}
    \label{eq:15}
    \begin{split}
    |\psi\rangle_{B}=&(x_{1}+x_{3}+ ... +x_{2^{N}-1})\prod^{N-1}_{i=1}|\delta_{kA_{i}}\rangle_{B_{i}}\otimes|0\rangle_{B_{N}}+(x_{2}+x_{4}+ ... +x_{2^{N}})\prod^{N-1}_{i=1}|\delta_{kA_{i}}\rangle_{B_{i}}\otimes|1\rangle_{B_{N}}\\
    =&|\phi\rangle_{A_{1}A_{2}...A_{N}},
    \end{split}
\end{equation}
and he has recovered Alice's original state;
    \item If the results contain even number of ``1", then Bob's N-particle system must be in the state
    \begin{equation}
    \label{eq:16}
    \begin{split}
    |\psi\rangle_{B}=&(x_{1}+x_{3}+ ...
    +x_{2^{N}-1})\prod^{N-1}_{i=1}|\delta_{kA_{i}}\rangle_{B_{i}}\otimes|0\rangle_{B_{N}}-(x_{2}+x_{4}+ ... +x_{2^{N}})\prod^{N-1}_{i=1}|\delta_{kA_{i}}\rangle_{B_{i}}\otimes|1\rangle_{B_{N}},
    \end{split}
    \end{equation}
\end{itemize}
and he only needs to execute the unitary transformation $\sigma_{z}$
on his particle $B_{N}$, and the state of $|\phi\rangle_{B}$ turns
into
\begin{equation}
\label{eq:17}
\begin{split}
(\sigma_{z})_{B_{N}}|\phi\rangle_{B}=&(\sigma_{z})_{B_{N}}[(x_{1}+x_{3}+
...
+x_{2^N-1})\prod^{N-1}_{i=1}|\delta_{kA_{i}}\rangle_{B_{i}}\otimes|0\rangle_{B_{N}}-(x_{2}+x_{4}+
...
+x_{2^N})\prod^{N-1}_{i=1}|\delta_{kA_{i}}\rangle_{B_{i}}\otimes|1\rangle
_{B{N}}]\\
=&(x_{1}+x_{3}+ ...
+x_{2^N-1})\prod^{N-1}_{i=1}|\delta_{kA_{i}}\rangle_{B_{i}}\otimes|0\rangle_{B_{N}}+(x_{2}+x_{4}+
...
+x_{2^N})\prod^{N-1}_{i=1}|\delta_{kA_{i}}\rangle_{B_{i}}\otimes|1\rangle
_{B{N}}\\
=&|\phi\rangle_{A_{1}A_{2}...A_{N}},
\end{split}
\end{equation}
and he has also recovered Alice's unknown state.
\end{itemize}
In short, by following our scheme above, MCQT of arbitrary
N-particle state can be successfully performed via ($N-1$) EPR pairs
and one (M+2)-particle GHZ state as quantum channel. The
correlations among Alice's measurement results, $Charlie_{i}$'s
unitary transformation, the number of ``1" in the agents'
measurement results and Bob's unitary transformations are shown in
Tab.~III.
\begin{table}[h]
\label{tab:3} \caption{Correlations among Alice's measurement
results, $Charlie_{i}$'s unitary transformation, the number of ``1"
in the agents' measurement results and Bob's unitary
transformation.}
\begin{center}
\begin{tabular}{c c c c c c} %% this creates two columns
%% |l|l| to left justify each column entry
%% |c|c| to center each column entry
%% use of \rule[]{}{} below opens up each row
\hline \hline
\rule[-1ex]{0pt}{3.5ex}  Alice's joint  & Alice's joint & $Charlie_{i}$'s unitary & Bob's unitary & Number & Bob's unitary\\
\rule[-1ex]{0pt}{3.5ex}  measurement result & measurement result & transformation & transformation & of & transformation\\
\rule[-1ex]{0pt}{3.5ex}  $|\psi\rangle_{A_{i}, D_{i}}$ & $|\psi\rangle_{A_{N}, D_{N}}$ & $U_{C_{j}}$ & $U_{i}$ & ``1" & $U_{N}$\\
\hline \hline %\rule[-1ex]{0pt}{3.5ex}
\multirow{4}{*}{$|\phi^{\pm}\rangle$} &
\multirow{2}{*}{$|\phi^{\pm}\rangle$} &
\multirow{2}{*}{$|0\rangle\langle0|+|1\rangle\langle1|$} &
\multirow{4}{*}{$|0\rangle\langle0|\pm|1\rangle\langle1|$}
& odd & $|0\rangle\langle0|+|1\rangle\langle1|$\\
& & & & even & $|0\rangle\langle0|-|1\rangle\langle1|$\\
& \multirow{2}{*}{$|\psi^{\pm}\rangle$} &
\multirow{2}{*}{$|0\rangle\langle1|\pm|1\rangle\langle0|$}
& & odd & $|0\rangle\langle0|+|1\rangle\langle1|$\\
& & & & even &
$|0\rangle\langle0|-|1\rangle\langle1|$\\
\multirow{4}{*}{$|\psi^{\pm}\rangle$} &
\multirow{2}{*}{$|\phi^{\pm}\rangle$} &
\multirow{2}{*}{$|0\rangle\langle0|+|1\rangle\langle1|$} &
\multirow{4}{*}{$|0\rangle\langle1|\pm|1\rangle\langle0|$}
& odd & $|0\rangle\langle0|+|1\rangle\langle1|$\\
& & & & even & $|0\rangle\langle0|-|1\rangle\langle1|$\\
& \multirow{2}{*}{$|\psi^{\pm}\rangle$} &
\multirow{2}{*}{$|0\rangle\langle1|\pm|1\rangle\langle0|$}
& & odd & $|0\rangle\langle0|+|1\rangle\langle1|$\\
& & & & even &
$|0\rangle\langle0|-|1\rangle\langle1|$\\
\hline
\end{tabular}
\end{center}
\end{table}
Note that each EPR pair used in the channel are in the state
$|\phi^{+}\rangle$, and the correlations of the cases for other EPR
pairs: $|\phi^{-}\rangle, |\psi^{+}\rangle$ and $|\psi^{-}\rangle$
are given in the appendix. In the next section, we give a MCQT
example to test the generality of our scheme by specifying $N$ and
$M$.

\section{Two-party controlled teleportation of arbitrary three-particle state: an example}
In this section, we test our general scheme by giving an example
where $N=3$ and $M=2$, which is a case of MCQT of arbitrary
tripartite state
\begin{equation}
|\phi\rangle_{A_{1}A_{2}A_{3}}=(x_{1}|000\rangle+x_{2}|001\rangle+x_{3}|010\rangle
+x_{4}|011\rangle+x_{5}|100\rangle+x_{6}|101\rangle+x_{7}|110\rangle+x_{8}|111\rangle).
\end{equation}
between Alice and Bob, jointly controlled by $Charlie_{1}$ and
$Charlie_{2}$.

Following our scheme in Sec. II, Alice firstly prepares two EPR
pairs in the state
$|\phi^{+}\rangle_{D_{i}B_{i}}=\frac{1}{\sqrt{2}}(|00\rangle+|11\rangle)_{D_{i}B_{i}}
(i=1, 2)$ and one four-particle GHZ state $|\phi^{+}\rangle_{
D_{3}B_{3}C_{1}C_{2}}=\frac{1}{\sqrt{2}}(|0000\rangle+|1111\rangle)_{D_{3}B_{3}C_{1}C_{2}}$.

Then Alice sends particles $B_{1}, B_{2}, B_{3}$ to Bob, and $C_{1}$
and $C_{2})$ to the two agents $Charlie_{1}$ and $Charlie_{2}$
respectively, keeping particles $(D_{1}, D_{2}, D_{3})$ to herself.
We have the whole system
\begin{equation}
\begin{split}
|\psi_{s}\rangle=&|\phi_{A_{1}A_{2}A_{3}}\rangle\otimes\frac{1}{\sqrt{2}}(|00\rangle+|11\rangle)_{D_{1}B_{1}}\otimes\frac{1}{\sqrt{2}}(|00\rangle+|11\rangle)_{D_{2}B_{2}}\\
&\otimes\frac{1}{\sqrt{2}}(|00\rangle+|11\rangle)_{D_{3}B_{3}}\otimes\frac{1}{\sqrt{2}}(|00000\rangle+|11111\rangle)_{D_{3}B_{3}C_{1}C_{2}}.
\end{split}
\end{equation}
Alice performs Bel1-joint measurements on pairs $(A_{1}, D_{1}),
(A_{2}, D_{2}), (A_{3}, D_{3})$ separately. Suppose the measurement
results are $|\phi^{+}\rangle, |\psi^{-}\rangle, |\phi^{-}\rangle$,
and the system turns to
\begin{equation}
\begin{split}
|\psi_{s}\rangle_{1}=&\langle\phi^{-}|\langle\psi^{-}|\langle\phi^{+}|\psi^{s}\rangle\\
=&(x_{1}|010\rangle-x_{3}|000\rangle+x_{5}|110\rangle-x_{7}|100\rangle)_{B_{1}B_{2}B_{3}}\otimes|00\rangle_{C_{1}C_{2}}\\
&-(x_{2}|011\rangle-x_{4}|001\rangle+x_{6}|111\rangle-x_{8}|101\rangle)_{B_{1}B_{2}B_{3}}\otimes|11\rangle_{C_{1}C_{2}}.
\end{split}
\end{equation}
According to Alice's measurement results, Bob performs
single-particle unitary transformations $I_{B_{1}},
(i\sigma_{y})_{B_{2}}, (\sigma_{z})_{B_{3}}$ on the particles
$B_{1}, B_{2}, B_{3}$ respectively, and the system state reads
\begin{equation}
\begin{split}
|\psi_{s}\rangle_{2}=&I_{B_{1}}(i\sigma_{y})_{B_{2}}(\sigma_{y})_{B_{3}}|\psi_{s}\rangle_{1}\\
=&(x_{1}|000\rangle+x_{3}|010\rangle+x_{5}|100\rangle+x_{7}|110\rangle)_{B_{1}B_{2}B_{3}}\otimes|00\rangle_{C_{1}C_{2}}\\
&+(x_{2}|001\rangle+x_{4}|011\rangle+x_{6}|101\rangle+x_{8}|111\rangle)_{B_{1}B_{2}B_{3}}\otimes|11\rangle_{C_{1}C_{2}}
\end{split}
\end{equation}
Now the two agents $Charlie_{1}$ and $Charlie_{2}$ perform
single-particle unitary operations $I_{z}$ on particle $C_{1},
C_{2}$ respectively according to Alice's measurement results on pair
$(A_{3}, D_{3})$. Then they respectively perform the Hadamard
transformations on particle $C_{1}, C_{2}$, and the whole system
reads
\begin{equation}
\begin{split}
|\psi_{s}\rangle_{3}=&\mathcal{H}_{C_{1}}\mathcal{H}_{C_{2}}|\psi_{s}\rangle_{2}\\
=&\frac{1}{2}(x_{1}|000\rangle+x_{3}|010\rangle+x_{5}|100\rangle+x_{7}|110\rangle)_{B_{1}B_{2}B_{3}}\otimes(|0\rangle+|1\rangle)_{C_{1}}\otimes(|0\rangle+|1\rangle)_{C_{2}}\\
&+\frac{1}{2}(x_{2}|001\rangle+x_{4}|011\rangle+x_{6}|101\rangle+x_{8}|111\rangle)_{B_{1}B_{2}B_{3}}\otimes(|0\rangle-|1\rangle)_{C_{1}}\otimes(|0\rangle-|1\rangle)_{C_{2}}\\
=&(x_{1}|000\rangle+x_{2}|001\rangle+x_{3}|010\rangle+x_{4}|011\rangle+x_{5}|100\rangle+x_{6}|101\rangle+x_{7}|110\rangle+x_{8}|111\rangle)_{B_{1}B_{2}B_{3}}\otimes(|00\rangle+|11\rangle)_{C_{1}C_{2}}\\
&+(x_{1}|000\rangle-x_{2}|001\rangle+x_{3}|010\rangle-x_{4}|011\rangle+x_{5}|100\rangle-x_{6}|101\rangle+x_{7}|110\rangle-x_{8}|111\rangle)_{B_{1}B_{2}B_{3}}\otimes(|01\rangle+|10\rangle)_{C_{1}C_{2}}
\end{split}
\end{equation}
Finally, $Charlie_{1}$ and $Charlie_{2}$ make single-particle
measurements on particles $C_{1}, C_{2}$ in basis
$\{|0\rangle,|1\rangle\}$ and publish their measurement results.

If the measurement results contain odd number of ``1", Bob's
N-particle system should be in the state
\begin{equation}
\label{eq:22}
 |\psi\rangle_{B}=(x_{1}|000\rangle+x_{2}|001\rangle+x_{3}|010\rangle+x_{4}|011\rangle+x_{5}|100\rangle+x_{6}|101\rangle+x_{7}|110\rangle+x_{8}|111\rangle)_{B_{1}B_{2}B_{3}},
\end{equation}
which is equivalent to the original unknown state of Alice;
otherwise, Bob's N-particle system reads
\begin{equation}
\label{eq:23}
 |\psi\rangle_{B}=(x_{1}|000\rangle-x_{2}|001\rangle+x_{3}|010\rangle-x_{4}|011\rangle+x_{5}|100\rangle-x_{6}|101\rangle+x_{7}|110\rangle-x_{8}|111\rangle)_{B_{1}B_{2}B_{3}},
\end{equation}
Bob then executes the unitary transformation $\sigma_{z}$ on
particle $B_{N}$, the state of $|\psi\rangle_{B}$ is thus recovered
to the original state of Alice.

In this example, Bob can fully recover the original state of Alice
with our general scheme, and it is clear that our scheme can be
specified into any other cases for $M\geq 0, N\geq 1$. Therefore, we
conclude that our theoretical MCQT scheme is highly general and can
be deterministically performed.
\section{Concluding remark}
In this paper, we propose a more general MCQT scheme of an arbitrary
N-particle state using $(N-1)$ EPR pairs and one $(M+2)$-particle
GHZ state together as quantum channel. We discuss the cases for each
kind of EPR pairs used in the quantum channel respectively and also
give a specified example to test the generality of our scheme.

We emphasize that, the QT schemes proposed in other literatures such
as Ref.~\cite{18,19,25} can be treated as special cases of our
scheme, and our scheme can also be used for quantum secure direct
communication (QSDC) and quantum secret sharing (QSS) tasks such as
the scheme in Ref.~\cite{221,222,24}. We hope this work will shed
some light for the prospective research on multiparty quantum
communications and quantum cryptography~\cite{29}.
\appendix
\acknowledgments Y.~Liu and J.~Ye are supported by the Innovation
Foundation of Aerospace Science \& Technology of the China Aerospace
Science \& Technology Corporation of China P. R. (Grant
No.~20060110, 2005-2007); Y.~Li is supported by the National
Undergraduate Novel Research Program from the Ministry of Education
of China P. R. (University Grant No.~0111181003, 2007-2008) and the
Undergraduate Innovation Foundation from Huazhong University of
Science \& Technology (2008-2009).
\section{The cases of using other EPR pairs in the quantum channel}
In the main context, we assume that the N-1 EPR pairs are in the
state $|\phi^{+}\rangle$. In fact, these EPR pairs can also be
either of the other EPR states $|\phi^{-}\rangle,|\psi^{+}\rangle$
or $|\psi^{-}\rangle$. The correlations among all the parameters of
each case are shown respectively in Tab.~IV,~V and~VI.
\begin{table}[h]
\label{tab:4} \caption{The case where N-1 pairs of
$|\phi^{-}\rangle$ are used in quantum channel.}
\begin{center}
\begin{tabular}{c c c c c c} %% this creates two columns
%% |l|l| to left justify each column entry
%% |c|c| to center each column entry
%% use of \rule[]{}{} below opens up each row
\hline \hline
\rule[-1ex]{0pt}{3.5ex}  Alice's joint  & Alice's joint & $Charlie_{i}$'s unitary & Bob's unitary & Number & Bob's unitary\\
\rule[-1ex]{0pt}{3.5ex}  measurement on & measurement on & transformation & transformation & of & transformation\\
\rule[-1ex]{0pt}{3.5ex}  $(A_{i}, D_{i})$ & $(A_{N}, D_{N})$ & $U_{C_{j}}$ & $U_{i}$ & ``1" & $U_{N}$\\
\hline \hline %\rule[-1ex]{0pt}{3.5ex}
\multirow{4}{*}{$|\phi^{\pm}\rangle$} &
\multirow{2}{*}{$|\phi^{\pm}\rangle$} &
\multirow{2}{*}{$|0\rangle\langle0|+|1\rangle\langle1|$} &
\multirow{4}{*}{$|0\rangle\langle0|\mp|1\rangle\langle1|$}
& odd & $|0\rangle\langle0|+|1\rangle\langle1|$\\
& & & & even & $|0\rangle\langle0|-|1\rangle\langle1|$\\
& \multirow{2}{*}{$|\psi^{\pm}\rangle$} &
\multirow{2}{*}{$|0\rangle\langle1|\pm|1\rangle\langle0|$}
& & odd & $|0\rangle\langle0|+|1\rangle\langle1|$\\
& & & & even &
$|0\rangle\langle0|-|1\rangle\langle1|$\\
\multirow{4}{*}{$|\psi^{\pm}\rangle$} &
\multirow{2}{*}{$|\phi^{\pm}\rangle$} &
\multirow{2}{*}{$|0\rangle\langle0|+|1\rangle\langle1|$} &
\multirow{4}{*}{$|0\rangle\langle1|\mp|1\rangle\langle0|$}
& odd & $|0\rangle\langle0|+|1\rangle\langle1|$\\
& & & & even & $|0\rangle\langle0|-|1\rangle\langle1|$\\
& \multirow{2}{*}{$|\psi^{\pm}\rangle$} &
\multirow{2}{*}{$|0\rangle\langle1|\pm|1\rangle\langle0|$}
& & odd & $|0\rangle\langle0|+|1\rangle\langle1|$\\
& & & & even &
$|0\rangle\langle0|-|1\rangle\langle1|$\\
\hline
\end{tabular}
\end{center}
\end{table}
\begin{table}[h]
\label{tab:5} \caption{The case where N-1 pairs of
$|\psi^{+}\rangle$ are used in quantum channel.}
\begin{center}
\begin{tabular}{c c c c c c} %% this creates two columns
%% |l|l| to left justify each column entry
%% |c|c| to center each column entry
%% use of \rule[]{}{} below opens up each row
\hline \hline
\rule[-1ex]{0pt}{3.5ex}  Alice's joint  & Alice's joint & $Charlie_{i}$'s unitary & Bob's unitary & Number & Bob's unitary\\
\rule[-1ex]{0pt}{3.5ex}  measurement on & measurement on & transformation & transformation & of & transformation\\
\rule[-1ex]{0pt}{3.5ex}  $(A_{i}, D_{i})$ & $(A_{N}, D_{N})$ & $U_{C_{j}}$ & $U_{i}$ & ``1" & $U_{N}$\\
\hline \hline %\rule[-1ex]{0pt}{3.5ex}
\multirow{4}{*}{$|\phi^{\pm}\rangle$} &
\multirow{2}{*}{$|\phi^{\pm}\rangle$} &
\multirow{2}{*}{$|0\rangle\langle0|+|1\rangle\langle1|$} &
\multirow{4}{*}{$|0\rangle\langle1|\pm|0\rangle\langle1|$}
& odd & $|0\rangle\langle0|+|1\rangle\langle1|$\\
& & & & even & $|0\rangle\langle0|-|1\rangle\langle1|$\\
& \multirow{2}{*}{$|\psi^{\pm}\rangle$} &
\multirow{2}{*}{$|0\rangle\langle1|\pm|1\rangle\langle0|$}
& & odd & $|0\rangle\langle0|+|1\rangle\langle1|$\\
& & & & even &
$|0\rangle\langle0|-|1\rangle\langle1|$\\
\multirow{4}{*}{$|\psi^{\pm}\rangle$} &
\multirow{2}{*}{$|\phi^{\pm}\rangle$} &
\multirow{2}{*}{$|0\rangle\langle0|+|1\rangle\langle1|$} &
\multirow{4}{*}{$|0\rangle\langle0|\pm|1\rangle\langle1|$}
& odd & $|0\rangle\langle0|+|1\rangle\langle1|$\\
& & & & even & $|0\rangle\langle0|-|1\rangle\langle1|$\\
& \multirow{2}{*}{$|\psi^{\pm}\rangle$} &
\multirow{2}{*}{$|0\rangle\langle1|\pm|1\rangle\langle0|$}
& & odd & $|0\rangle\langle0|+|1\rangle\langle1|$\\
& & & & even &
$|0\rangle\langle0|-|1\rangle\langle1|$\\
\hline
\end{tabular}
\end{center}
\end{table}
\begin{table}[h]
\label{tab:6} \caption{The case where N-1 pairs of
$|\psi^{-}\rangle$ are used in quantum channel.}
\begin{center}
\begin{tabular}{c c c c c c} %% this creates two columns
%% |l|l| to left justify each column entry
%% |c|c| to center each column entry
%% use of \rule[]{}{} below opens up each row
\hline \hline
\rule[-1ex]{0pt}{3.5ex}  Alice's joint  & Alice's joint & $Charlie_{i}$'s unitary & Bob's unitary & Number & Bob's unitary\\
\rule[-1ex]{0pt}{3.5ex}  measurement on & measurement on & transformation & transformation & of & transformation\\
\rule[-1ex]{0pt}{3.5ex}  $(A_{i}, D_{i})$ & $(A_{N}, D_{N})$ & $U_{C_{j}}$ & $U_{i}$ & ``1" & $U_{N}$\\
\hline \hline %\rule[-1ex]{0pt}{3.5ex}
\multirow{4}{*}{$|\phi^{\pm}\rangle$} &
\multirow{2}{*}{$|\phi^{\pm}\rangle$} &
\multirow{2}{*}{$|0\rangle\langle0|+|1\rangle\langle1|$} &
\multirow{4}{*}{$|0\rangle\langle1|\mp|0\rangle\langle1|$}
& odd & $|0\rangle\langle0|+|1\rangle\langle1|$\\
& & & & even & $|0\rangle\langle0|-|1\rangle\langle1|$\\
& \multirow{2}{*}{$|\psi^{\pm}\rangle$} &
\multirow{2}{*}{$|0\rangle\langle1|\pm|1\rangle\langle0|$}
& & odd & $|0\rangle\langle0|+|1\rangle\langle1|$\\
& & & & even &
$|0\rangle\langle0|-|1\rangle\langle1|$\\
\multirow{4}{*}{$|\psi^{\pm}\rangle$} &
\multirow{2}{*}{$|\phi^{\pm}\rangle$} &
\multirow{2}{*}{$|0\rangle\langle0|+|1\rangle\langle1|$} &
\multirow{4}{*}{$|0\rangle\langle0|\mp|1\rangle\langle1|$}
& odd & $|0\rangle\langle0|+|1\rangle\langle1|$\\
& & & & even & $|0\rangle\langle0|-|1\rangle\langle1|$\\
& \multirow{2}{*}{$|\psi^{\pm}\rangle$} &
\multirow{2}{*}{$|0\rangle\langle1|\pm|1\rangle\langle0|$}
& & odd & $|0\rangle\langle0|+|1\rangle\langle1|$\\
& & & & even &
$|0\rangle\langle0|-|1\rangle\langle1|$\\
\hline
\end{tabular}
\end{center}
\end{table}

\end{document}